\documentclass[a4paper,11pt]{article}
\pdfoutput=1 

\usepackage{jheppub} 

\usepackage[T1]{fontenc} 
\usepackage{natbib,amsmath,amssymb,braket}
\usepackage{enumitem}

\newtheorem{definition}{Definition}[section]

\newtheorem{assertion}{Assertion}[section]
\newtheorem{proposition}{Proposition}[section]

\title{Notes on algebraic perspective on neural network conformal theories}

\author[a,1]{Samuel Leutheusser}
\author[a,b,c,d,2]{Joydeep Naskar}

\affiliation[a]{Kavli Institute for Theoretical Physics, Santa Barbara, CA 93106, USA.}
\affiliation[b]{Department of Physics, Northeastern University, Boston, MA, 02115, USA}
\affiliation[c]{The NSF AI Institute for Artificial Intelligence and Fundamental Interactions, Cambridge, MA, USA.}
\affiliation[d]{Beijing Institute of Mathematical Sciences and Applications, Huaibei Town, Huairou District, Beijing 101408, China}

\emailAdd{sam.leutheusser@gmail.com}
\emailAdd{naskar.j@northeastern.edu}

\abstract{Neural network conformal theories were first introduced in \cite{Halverson:2024axc} and exact two-, three- and four-point correlators were computed from an ensemble of infinite-width neural networks. In particular, the four-point correlator of any scalar operator satisfied crossing-symmetry and conformal block decomposition across different channels were in agreement without any constraint on moments. An infinite tower of degenerate operators with the same scaling dimension were found. In this note, we take a closer look at the structure of correlators and relation between operators. We also layout a recipe to construct the algebra of the theory.}

\begin{document} 
\maketitle
\flushbottom

\section{Introduction}
\label{sec:intro}
The neural network-field theory correspondence \cite{Halverson:2020trp, Halverson:2021aot,Halverson:2024hax} and its associated universal approximation theorem \cite{Ferko:2026axm} suggests that any Euclidean quantum field theory admits a neural network description with a countably infinite number of parameters. The art of constructing (quantum) fields from the outputs of neurons and computing exact correlators using the parameter space description, exploiting its relation with function space \cite{Maiti:2021fpy} has recently gathered interest and several interesting works appeared in this direction \cite{Frank:2026bui,Frank:2025zuk,Ferko:2025ogz,Halverson:2024axc,Demirtas:2023fir,Halverson:2021aot,Capuozzo:2025ozt,Robinson:2025ybg,Ageev:2026ofv,Sen:2025vzl,Howard:2024kfd,Huang:2025ipy,Ferko:2026ken,Ferko:2026kkm,Dogra:2026hfa,Jefferson:2026nio,Ferko:2026ukw,Harvey:2026iqd}\footnote{See \cite{Balassa:2025bgt,Balassa:2025gwh,Balassa:2025biq,Martyn:2022oll,Lee:2025mti,Ageev:2026qyh,Hashimoto:2024aga,Halverson:2026pmb,Balassa:2026lil} for some other interesting works on machine learning approaches to quantum field theory.}. This gives us a non-Lagrangian approach to study (quantum) field theory, where the correlators are computed using the statistics of neural-network parameters. In this probabilistic approach to quantum field theory, the two crucial ingredients are therefore, engineering symmetries of the correlators and cooking the dynamics of the theory through an appropriate choice of the joint probability distribution. The interplay between symmetries of correlators and neural network architecture  are somewhat well-studied, but the understanding of probability distribution function and the dynamics of an interacting theory are less understood. Moreover, the computation of correlators in these theories mimic the path-integral techniques, the discussion of Hilbert space of these theories have not been discussed.

In this paper, we will focus on the conformal theories built using the embedding formalism. One interesting class of theories constructed are the Euclidean non-unitary generalized free scalar theories with conformal symmetries \cite{Halverson:2024axc} with non-positive integer scaling dimensions\footnote{Note that we will limit our discussion to this particular class of theories, whereas the construction proposed in \cite{Halverson:2024axc} can be more general.}. In the four-dimensional toy-model theory considered in \cite{Halverson:2024axc}, one takes a neural conformal field $\phi(x)$ with scaling dimension $\Delta=-1$ and compute its two- and four-point functions, both of which respect conformal symmetry. The four-point function respects crossing symmetry and one can perform a conformal block expansion to get the spectrum of operators in the $\phi\times\phi$ neural-operator-product-expansion (nOPE)\footnote{We call it a neural-operator-product-expansion (nOPE), because it behaves like an `OPE', however, not at an operator level, but after taking expectation values.}, which asserts the existence of neural conformal fields of dimension $0$ and $-2$. One might as well have started with a neural field $\mathcal{O}_n$ of any scaling dimension $\Delta=-n$ (where $n\in \mathbb{Z}_+$) and compute the correlators, including mixed ones and extract respective operator spectra. In summary, all the neural-field correlators respect conformal symmetry. However, we find an infinite number of degenerate neural-fields at any fixed scaling dimension. It was unclear if this degeneracy is artificial. This can be resolved by understanding the Hilbert space of the theory.

Another feature of the theory was that the four-point functions didn't constrain the moments of the parameter distribution function, despite non-trivial constraints from conformal bootstrap. However, the consistency of mixed correlators with bootstrap conditions required contributions from the off-diagonal terms in the Gram matrix\footnote{This point was not discussed properly in \cite{Halverson:2024axc}, so will discuss it in this paper.}. Naively, this gives an impression that any probability distribution satisfying rotational invariance sufficed to give conformal theories with infinitely degenerate infinite number of neural fields. It turns out that the degeneracy of operators and constraints between moments are deeply interlinked. In particular, one needs a positive definite Gram matrix to construct the Hilbert space and the moments are constrained non-trivially. However, even in the absence of Euclidean reflection positivity, we can define Krein-spaces under some assumptions. Nevertheless, we will not use any of those to construct our algebra, i.e, we will avoid going in the Haag-Kastler formalism, and instead use the conformal group and its generators to define the algebra. In doing so, we are only assuming the state-operator correspondence. It is however, tempting to prove those assumptions for Krein positivity and take a Haag-Kastler route to define the algebra for these theories.

\subsection*{Summary and organization}
The organization of this note is as follows. In section \ref{sec:nn-ct}, we review the construction of neural network conformal theories and discuss the non-unitary scalar theories. In section \ref{sec:moments}, we do some numerics to illuminate on the parameter-space Gram matrix at fixed Euclidean separation. Section \ref{sec:algebra-correlators} contains our central result on the construction of a $*$-algebra for our neural network conformal theories. We conclude with some disucssions in section \ref{sec:discuss}.

\section{Neural network conformal theories}
\label{sec:nn-ct}
\subsection{A Brief Review of the Construction}
\label{subsec:review}
Consider a $(D+2)$-dimensional Euclidean theory with $SO(D+2)$ invariant correlators. We will pick a $SO(D+2)$ invariant joint probability distribution function $P(\Theta)$\footnote{Note that $\Theta$ is not a single parameter but a label for the collection of all parameters.} and a homogeneous neural network field $\Phi(X)$\footnote{Strictly, it is $\Phi_{\Theta}(X)$ but we will suppress the subscript $\Theta$.} satisfying,
\begin{equation}\label{eq:homo}
    \Phi(\lambda X)= \lambda^{-\Delta}\Phi(X),
\end{equation}
whose parameters are drawn from $P(\Theta)$. The partition function for this theory can be written as
\begin{equation}\label{eq:PF}
    Z[J]=\int D\Theta P(\Theta)  e^{\int d^{D+2}X J(X) \Phi(X)},
\end{equation}
and expectation values (say of some $\mathcal{O}$)  can be computed by
\begin{equation}\label{eq:exp}
    \langle \mathcal{O} \rangle = \int d\Theta P(\Theta) \mathcal{O}.
\end{equation} 
In particular, we are interested in computing the $n$-point function (say, of $\Phi$), which can be computed by
\begin{equation}\label{eq:n-pt-e}
    G_E^{(n)}(X_1, \cdots , X_n) =  \int d\Theta P(\Theta) \Phi(X_1) \cdots \Phi(X_n),
\end{equation}
where the subscript $E$ denotes Euclidean and $X$s are ($D+2$)-dimensional Euclidean position vectors. Now we will Wick rotate the correlator to Minkowski space, giving us $G_M^{(n)}(X_1, \cdots , X_n)$ which we further restrict to the projective null cone, i.e., schematically, we have
\begin{equation}\label{eq:n-pt-m}
    G_E^{(n)}(X_1, \cdots , X_n) \xrightarrow{\text{Wick}} G_M^{(n)}(X_1, \cdots , X_n)  \xrightarrow{\text{PNC}} G_{PNC}^{(n)}(X_1, \cdots , X_n)
\end{equation}
where the subscripts $M$ and $PNC$ denotes Minkowski and projective null cone respectively.
We will use the standard technology of the embedding formalism of CFT (see \cite{Rychkov:2016iqz} for a review). As we restrict ourselves to the Poincare section of the projective null cone, it leaves us with correlators of conformal fields $\phi$ in $D$-dimensions, i.e we identify the following,
\begin{equation}\label{eq:pnc}
    \begin{aligned}
        & \Phi(X)_{PNC}\equiv \phi(x),\\
        & G_{PNC}^{(n)}(X_1, \cdots , X_n) \equiv G^{(n)}(x_1, \cdots , x_n),
    \end{aligned}
\end{equation}
where $G^{(n)}(x_1, \cdots , x_n)$ is the $n$-point correlator of the conformal field $\phi$, and the coordinates $x$ and $X$ are related by,
\begin{equation}\label{eq:coords}
    X_{\mu}= \left(\frac{1+x^2}{2}, x_{a}, \frac{1-x^2}{2} \right),
\end{equation}
where $a$ runs from $1$ to $D$, and $\mu$ runs from $0$ to $D+1$.

\subsection{The Non-unitary Scalar Conformal Theory}
We will now focus on a particular class of theory from \cite{Halverson:2024axc}, namely the non-unitary scalar conformal theory. In this theory, the conformal fields have the form
\begin{equation}\label{eq:mn-fields-def}
    \mathcal{O}_{m,n}(X)= \left (\theta \cdot \theta\right)^m \left(\theta \cdot X\right)^n,
\end{equation}
where $\theta$s are drawn from a $SO(D+2)$ invariant, independent and identical distribution $P(\theta)$ and $\Delta=-n$ is the scaling dimension, with the simplest example being
\begin{equation}\label{eq:phi-def}
    \Phi(X)\equiv \mathcal{O}_{0,1}(X) = \left(\theta \cdot X\right),
\end{equation}
whose two-, three- and four-point correlators are given by,
\begin{equation}\label{eq:phi-2-4}
\begin{aligned}
    & G^{(2)}_{{0,1};{0,1}}(x_1,x_2) = \mu_2 x_{12}^2, \\
    & G^{(3)}_{{0,1};{0,1};{0,1}}(x_1,x_2,x_3) = 0\\
    & G^{(4)}_{{0,1};{0,1};{0,1};{0,1}}(x_1,x_2,x_3,x_4)= \frac{\mu_4}{3} \left(x_{12}^2 x_{34}^2+ x_{13}^2 x_{24}^2+ x_{14}^2 x_{23}^2 \right)= g(u,v) x_{12}^2 x_{34}^2 ,
\end{aligned}
\end{equation}
where $x_{12}^2=|x_1-x_2|^2$, $u=\frac{x_{12}^2x_{34}^2}{x_{13}^2x_{24}^2}$ and $v=\frac{x_{14}^2x_{23}^2}{x_{13}^2x_{24}^2}$ are the cross-ratios, and 
\begin{equation}
    g(u,v)=\frac{\mu_4}{3} \left(1+\frac{1}{u}+\frac{v}{u}\right).
\end{equation}

Now let us restrict ourselves to $D=4$. We can write a conformal block decomposition of the four-point function in \ref{eq:phi-2-4} as follows,
\begin{equation}\label{eq:phi_guv}
    g(z,\Bar{z})=\frac{\mu_4}{3}\left(2 \mathbf{g}_{-2}(z,\Bar{z}) +\frac{4}{3}\mathbf{g}_{0}(z,\Bar{z})\right)
\end{equation}
where $\mathbf{g}_{\Delta}$ is the four-dimensional conformal block associated with a scalar operator of scaling dimension $\Delta$\cite{Dolan:2000ut}, which in this case can be written as,
\begin{equation}
    \begin{split}
        & \mathbf{g}_{0}(z,\Bar{z})= 1, \\
        & \mathbf{g}_{-2}(z,\Bar{z})= \frac{1}{z\Bar{z}}-\frac{1}{2z}-\frac{1}{2\Bar{z}}+\frac{1}{3},
    \end{split}
\end{equation}
and the variables $(z,\Bar{z})$ are related to $(u,v)$ by the following relation,
\begin{equation}\label{eq:zzbar-uv}
    u=z\Bar{z}, \quad v=(1-z)(1-\Bar{z}).
\end{equation}
It can be explicitly shown that these contributions come from the following operators,
\begin{equation}
    \begin{split}
        & \Delta=0: \quad \mathcal{O}_{1,0} = (\theta \cdot \theta), \\
        & \Delta=-2: \quad \mathcal{O}_{0,2} = (\theta\cdot X)^2.
    \end{split}
\end{equation}
One can similarly compute correlation functions of all other fields (\ref{eq:mn-fields-def}) and extract operator spectra. The general formula for two-point functions of arbitrary operators $\mathcal{O}_{m,n}$ and $\mathcal{O}_{m',n'}$ is given by,
\begin{equation}\label{eq:2-pt-arbitrary}
    G^{(2)}_{{m,n};{m',n'}}(x_1,x_2) = \delta_{nn'} \frac{\mu_{2k}}{(2k-1)!!} \cdot n! \cdot R_p(D+2+2n) \cdot x_{12}^{2n},
\end{equation}
with $p = m + m'$ and $k = p + n$ and we define the \emph{Rising factorial} as,
\begin{equation}
R_p(a) = a(a+2)(a+4)\cdots(a+2(p-1)) = \prod_{j=0}^{p-1}(a + 2j),
\end{equation}
with the convention $R_0(a) = 1$. See Appendix \ref{app:a} for a derivation of this formula.

\subsubsection{The puzzle of mixed four-point functions}
We will recall on puzzle regarding the mixed four-point correlators encountered in \cite{Halverson:2024axc}. For example, consider the correlator
\begin{equation}\label{eq:mixed-phi-phi2}
    G_{0,1;0,1;0,2;0,2}(x_1,x_2,x_3,x_4) = \frac{2\mu_6}{15} (x_{12})^2 (x_{34})^4 \left[1+2\left(\frac{1}{u}+ \frac{v}{u}\right)\right]
\end{equation}
The conformal block decomposition of $g(u,v)$ is given by,
\begin{equation}
    \label{eq:gmixed_cbd}
    g(z,\Bar{z})= \frac{2\mu_6}{15} \left(4 \mathbf{g}_{-2}(z,\Bar{z}) + \frac{5}{3} \mathbf{g}_{0}(z,\Bar{z}) \right),
\end{equation}
which is consistent with the expectation that $G_{0,1;0,1;0,2;0,2}$ can contain only those operators that are present in the spectra of \emph{both} $\mathcal{O}_{0,1}\times\mathcal{O}_{0,1}$ nOPE and $\mathcal{O}_{0,2}\times\mathcal{O}_{0,2}$ nOPE, which in this case are the dimension $0$ and dimension $(-2)$ fields. However, note that while on one hand, the $\mathcal{O}_{0,1}\times\mathcal{O}_{0,1}$ nOPE contains the neural fields $\mathcal{O}_{1,0}$ and $\mathcal{O}_{0,2}$, on the other hand, $\mathcal{O}_{0,2}\times\mathcal{O}_{0,2}$ nOPE contains the neural fields $\mathcal{O}_{2,0}$ and $\mathcal{O}_{1,2}$. However, since the correlators $G^{(2)}_{1,0;2,0}\neq0$ and $G^{(2)}_{0,2;1,2}\neq0$, a simple contraction with appropriate normalization matches the answer given in eq \ref{eq:gmixed_cbd}. More specifically, we have
\begin{equation}
\begin{split}
    & G^{(3)}_{0,1;0,1;1,0}(X,Y,Z)=\frac{\mu_4}{3}(D+4) (X\cdot Y) = \frac{8\mu_4}{3} (X\cdot Y)\\
    & G^{(3)}_{0,2;0,2;2,0}(X,Y,Z)=\frac{\mu_8}{105}(2(D+2)^2 + 20(D+2) + 48) (X\cdot Y)^2=\frac{240 \mu_8}{105} (X\cdot Y)^2\\
    & G^{(2)}_{{1,0};{1,0}}(X,Y)=  \frac{\mu_4}{3}((D+2)^2 + 2(D+2))=\frac{48 \mu_4}{3}=c_{\mathcal{O}_{1,0} \mathcal{O}_{1,0}}\\
    & G^{(2)}_{{2,0};{2,0}}(X,Y) = \frac{\mu_8}{105}((D+2)^4 + 12 (D+2)^3 + 44 (D+2)^2 + 48 (D+2)) = \frac{5760 \mu_8}{105}=c_{\mathcal{O}_{2,0}\mathcal{O}_{2,0}}\\
    &  G^{(2)}_{{2,0};{1,0}}(X,Y)= \frac{\mu_6}{15} ((D+2)^3+6(D+2)^2+8(D+2))=\frac{480 \mu_6}{15}=c_{\mathcal{O}_{2,0} \mathcal{O}_{1,0}}.
\end{split}
\end{equation}
The coefficient of $\mathbf{g}_{0}$ in conformal block decomposition \ref{eq:gmixed_cbd} is given by
\begin{equation}\label{eq:1122_0}
    \lambda_{\Phi\Phi\mathcal{O}_{1,0}} \lambda_{\Phi^2\Phi^2\mathcal{O}_{2,0}} \left[\frac{c_{\mathcal{O}_{2,0}\mathcal{O}_{1,0}}}{c_{\mathcal{O}_{1,0} \mathcal{O}_{1,0}}c_{\mathcal{O}_{2,0}\mathcal{O}_{2,0}}}\right]=\frac{8\mu_4}{3}\frac{240\mu_8}{105} \frac{480\mu_6}{15}\frac{105}{5760\mu_8}\frac{3}{48\mu_4}=\frac{2\mu_6}{9}.
\end{equation}
Does this prescription really work? Let us consider the correlators
\begin{equation}
    \begin{split}
        & G^{(3)}_{0,1;0,1,0,2}(X,Y,Z)=\frac{2\mu_4}{3}(X\cdot Z)(Y\cdot Z)\\
        & G^{(3)}_{0,2;0,2;{1,2}}(X,Y,Z) = \frac{\mu_8}{105} (8(D+2) + 48) (X\cdot Y) (X\cdot Z) (Y\cdot Z) =\frac{96 \mu_8}{105} (X\cdot Y) (X\cdot Z) (Y\cdot Z)\\
        & G^{(2)}_{{1,2} {1,2}}(X,Y)= \frac{\mu_8}{105} (2(D+2)^2 + 20(D+2) + 48) (X\cdot Y)^2 = \frac{240\mu_8}{105}(X\cdot Y)^2= c_{\mathcal{O}_{1,2}} \mathcal{O}_{1,2} (X\cdot Y)^2\\
        & G^{(2)}_{0,2;0,2}(X,Y)=\frac{2\mu_4}{3} (X\cdot Y)^2 = c_{\Phi^2\Phi^2} (X\cdot Y)^2\\
        & G^{(2)}_{0,2;{1,2}}=\frac{\mu_6}{15}(2(D+2)+8)(X\cdot Y)^2 =\frac{20\mu_6}{15} (X\cdot Y)^2 = c_{\Phi^2\mathcal{O}_{1,2}}(X\cdot Y)^2 
    \end{split}
\end{equation}
The coefficient of $\mathbf{g}_{-2}$ is given by:
\begin{equation}\label{eq:1122_-2}
    \lambda_{\Phi\Phi\mathcal{O}_{0,2}}\lambda_{\Phi^2\Phi^2\mathcal{O}_{1,2}} \left[ \frac{c_{\Phi^2\mathcal{O}_{1,2}}}{c_{\Phi^2\Phi^2}c_{\mathcal{O}_{1,2}\mathcal{O}_{1,2}}} \right]=\frac{8\mu_6}{15}
\end{equation}
Thus we see that~(\ref{eq:1122_0}) and~(\ref{eq:1122_-2}) do match the CBD coefficients in~(\ref{eq:gmixed_cbd}).

So where is the puzzle? The puzzle is that this is not how you compute correlators for degenerate operators in a standard conformal field theory. In practice, you need a Gram matrix $\mathcal{G}^{IJ}$, where
\begin{equation}
    \mathcal{G}^{IJ}\mathcal{G}_{JK}= \delta^I_K, \quad \text{and} \quad \mathcal{G}_{IJ}=G^{(2)}_{\mathcal{O_I} \mathcal{O_J}},
\end{equation}
where we have assumed the invertibility of Gram matrix $\mathcal{G}_{IJ}$, i.e., det$(\mathcal{G})\neq 0$. More precisely, since the OPE is simply the insertion of the identity in radial quantization one should write the coefficient of $\mathbf{g}_{0}$ (with appropriate normalization) as
\begin{equation}
    G^{IJ} \lambda_{\mathcal{O}_{0,1}\mathcal{O}_{0,1}\mathcal{O}_I} \lambda_{\mathcal{O}_{0,2}\mathcal{O}_{0,2}\mathcal{O}_J}.
\end{equation}
Since we will need the information about an infinite number of operators to find the inverse matrix $G^{IJ}$, this was never attempted in \cite{Halverson:2024axc}. If we take the Gram-submatrix with low-lying operators, clearly the answer doesn't match the coefficient found by conformal block decomposition.

\subsubsection*{Comments on this puzzle}
Before we attempt to resolve this puzzle, let us clarify some aspects about the nOPE.
We should not identify nOPE as an operator relation, such as done in an OPE, and instead treat it like what it actually is, i.e., a relation that holds only in expectation value, i.e.,
\begin{equation}\label{eq:ope-op}
    \mathcal{O}_{m_1,n_1}\mathcal{O}_{m_2,n_2}\neq \sum_{p,q} \lambda_{\mathcal{O}_{m_1,n_1}\mathcal{O}_{m_2,n_2}\mathcal{O}_{p,q}}
\end{equation}
and, instead, the correct understanding is,
\begin{equation}\label{eq:ope-exp}
    \langle \mathcal{O}_{m_1,n_1}\mathcal{O}_{m_2,n_2} \mathcal{O}_{m_3,n_3}\mathcal{O}_{m_4,n_4}\rangle = \sum_{P,Q,R,S} \lambda_{\mathcal{O}_{m_1,n_1}\mathcal{O}_{m_2,n_2}\mathcal{O}_{P,Q}}
    \lambda_{\mathcal{O}_{m_3,n_3}\mathcal{O}_{m_4,n_4}\mathcal{O}_{R,S}} \frac{\langle \mathcal{O}_{P,Q} \mathcal{O}_{R,S} \rangle}{\sqrt{\langle \mathcal{O}_{P,Q} \mathcal{O}_{P,Q} \rangle \langle \mathcal{O}_{R,S} \mathcal{O}_{R,S} \rangle}},
\end{equation}
where we have suppressed the space-time points. In philosophy, the emergence of nOPE is through the averaging and it can connect degenerate operators with different $\theta$-homogeneity. Therefore, since we cannot interpret the nOPE as an operator relation. So the next question is, if it is not an operator relation, should we not treat it as an expansion of identity operator? Well, definitely not in the standard sense. However, there is an analogue of it as an expansion around dimension $0$ operators $\mathcal{O}_{M,0}$ for different $M$ as we explain below.

Consider the mixed four-point function
$\langle \mathcal{O}_{m_1,n_1}\mathcal{O}_{m_2,n_2} \mathcal{O}_{m_3,n_3}\mathcal{O}_{m_4,n_4}\rangle$ in the $s$-channel.
\begin{equation}
    \begin{aligned}
        \langle \mathcal{O}_{m_1,n_1}\mathcal{O}_{m_2,n_2} \mathcal{O}_{m_3,n_3}\mathcal{O}_{m_4,n_4}\rangle & = \langle \mathcal{O}_{m_1,n_1}\mathcal{O}_{m_2,n_2}\textbf{1}_{d_{12}}\textbf{1}_{d_{34}} \mathcal{O}_{m_3,n_3}\mathcal{O}_{m_4,n_4}\rangle \\
        & = \langle \mathcal{O}_{m_1,n_1}\mathcal{O}_{m_2,n_2} \sum_{m,n}\frac{\ket{\mathcal{O}_{m,n}}\bra{\mathcal{O}_{m,n}}}{\sqrt{\langle \mathcal{O}_{m,n} |\mathcal{O}_{m,n}\rangle}} \sum_{m',n'}\frac{\ket{\mathcal{O}_{m',n'}}\bra{\mathcal{O}_{m',n'}}}{\sqrt{\langle \mathcal{O}_{m,n} |\mathcal{O}_{m,n}\rangle}} \mathcal{O}_{m_3,n_3}\mathcal{O}_{m_4,n_4}\rangle \\
        & = \frac{1}{\mathcal{K}}\sum_{m,m',n} \langle \mathcal{O}_{m_1,n_1}\mathcal{O}_{m_2,n_2} \ket{\mathcal{O}_{m',n}} \langle \mathcal{O}_{m,n} | \mathcal{O}_{m',n} \rangle \bra{\mathcal{O}_{m',n}} \mathcal{O}_{m_3,n_3}\mathcal{O}_{m_4,n_4}\rangle,
    \end{aligned}
\end{equation}
where in the first equality we expanded on two complete sets of operators of $\theta$-degree $d_{12}$ and $d_{34}$ respectively, which are given by,
\begin{equation}
\begin{split}
    & d_{12}=2(m_1+m_2)+(n_1+n_2) \\
    & d_{34}=2(m_3+m_4)+(n_3+n_4),
\end{split}  
\end{equation}
and in the second equality we have used the relation that $\langle \mathcal{O}_{m,n} | \mathcal{O}_{m',n'}\rangle \propto \delta_{nn'}$, and have absorbed some scalar numbers into the the number $\mathcal{K}$. This structure tells us that the two nOPE channels comes from performing an identity-like expansion across two different dimension-$0$ operators, where the sum over operators are over $\theta$-degree preserving sectors. However, this $\theta$-degree is not a global symmetry and operators in two different $\theta$-degree sectors can have non-trivial two-point functions. To illustrate some examples, it is easy to see that for the case given eq. \ref{eq:1122_0}, we have $m=1,m'=2, n=0$ and for the case given in eq. \ref{eq:1122_-2}, we have $m=0,m'=1,n=2$. More generally, one can write down the nOPE as a $\theta$-sector preserving expansion, imagined as,
\begin{equation}\label{eq:eq_nOPE_imagine}
    \mathcal{O}_{m_1,n_1} \times \mathcal{O}_{m_2,n_2} \sim \sum_{p,q} \mathcal{O}_{p,q},
\end{equation}
where $2p+q=2(m_1+m_2)+(n_1+n_2)$. The number of operators in a $\theta$-degree sector $S$ grows comparatively as fast as the number of partitions of $S$. Note that however, one must not take \ref{eq:eq_nOPE_imagine} as an operator relation, and remember eq. \ref{eq:ope-op}. We reiterate that these characteristics are unusual in standard CFTs and this makes the algebraic structure of these NN-CTs different. While one may find it tempting to think of different dimension $0$ operators $\mathcal{O}_{m,0}$ as degenerate vacuua, and the operators $\mathcal{O}_{m,n}$ as excitations around these vacuua and the non-zero overlaps between different vacuua (and their excitations) as tunneling between different vacuua sectors of the theory, however, this interpretation is not well-motivated, as explained in the following subsubsection. On a separate note, since the four-point functions play a prominent role for expectation values to hold in \ref{eq:ope-exp}, it would be nice to explore if this system satisfies the weakened definition of conformal correlator systems \cite{Ribault:2026bbu}.

\subsubsection{Map between primaries through descendants}
In the embedding space, one can generate the descendants of a scalar primary operator of dimensions $\Delta$ by applying the Thomas-Todorov operator $\mathbf{D}_A$ \cite{Dobrev:1975ru,Fortin:2016dlj} written as,
\begin{equation}\label{eq:todorov_op}
    \mathbf{D}_A= \left(\frac{D}{2} + X \cdot \partial_X\right)\,\partial_A - \frac{1}{2} X_A\, \partial^2 ,
\end{equation}
which generates a spin-1 operator of dimension $\Delta+1$. In our case, we are interested in scalar operators\footnote{One may study these spinning operators as well, however, we do not partake that task in this paper. See \cite{Dogra:2026hfa} for a neural-network field theory construction spinning conformal correlators.}, so we will act with the Todorov operator twice. It turns out that,

\begin{equation}
    \mathbf{D}^2 \mathcal{O}_{m,n} \equiv\mathbf{D}^A\mathbf{D}_A \mathcal{O}_{m,n} \propto (X\cdot X) \mathcal{O}_{m+1,n-2}=0,
\end{equation}
since the factor $X\cdot X$ is zero. Therefore, the operator $\mathcal{O}_{m+1,n-2}$ is not a genuine descendant of $\mathcal{O}_{m,n}$, well-defined on the null cone. These two operators are related by a map that is close to a descendant, however, not quite a descendant, at least, not at the level of a well-defined field on the null cone. Interestingly, their correlators do not care about well-definedness on the null cone, and this subtlety can be ignored at the level of correlators.

Let us modify the prescription above by acting with a Todorov operator followed by an partial derivative $\partial_A$, i.e, $\Box\equiv \mathbf{D}^A \partial_A$, giving us a scalar primary of dimension $\Delta+2$.
\begin{equation}\label{eq:todorov_twice}
    \Box \mathcal{O}_{m,n} \equiv\mathbf{D}^A \partial_A \mathcal{O}_{m,n}= n(n-1) \left(\frac{D}{2}+\frac{n}{2}-1 \right) \mathcal{O}_{m+1,n-2},
\end{equation}
thereby showing that all primaries $\mathcal{O}_{m,n}$ with $m\neq 0$ are also \emph{quasi-descendants} of the operator $\mathcal{O}_{0,2m+n}$. In other words, the primary operators $\mathcal{O}_{0,S}$ are the \emph{root} primary operator that are not descendants of any other primaries and all other primaries appearing with the same $\theta$-degree $S$ are its quasi-descendants.

This interpretations leads us to the OPE relation between primaries,
\begin{equation}
    \mathcal{O}_{0,n_1} \times \mathcal{O}_{0,n_2} \sim \left[\mathcal{O}_{0,n_1+n_2}+ \text{descendants} \right],
\end{equation}
where the primary part $\mathcal{O}_{0,n_1+n_2}$ is an operator relation and the quasi-descendants appear only in expectation values.

One should note however, these quasi-descendants are not genuine descendants, and therefore all operators $\mathcal{O}_{m,n}$ are primaries.

\subsection{A Quick Review of Large-$N$ Limit of Neural Network Ensembles}\label{subsec:larg-N}
We can also take an ensemble of $N$ neural network fields $\Phi(X)$ to construct a field $\varphi(X)$,
\begin{equation}\label{eq:N-field-ensemble}
    \varphi(X)=\frac{1}{\sqrt{N}}\sum_{i=1}w_i\Phi_i(X),
\end{equation}
where $\{w_i\}\sim P(w)$ i.i.d and $\Phi_i(X)$ is drawn from a collection of fields defined by \ref{eq:phi-def}(or, more generally, any other field defined by \ref{eq:mn-fields-def}). Let us assume the following properties for the distribution of each $w_i$:
\begin{equation}\label{eq:w-moment-conditions}
    \langle w_i^{k+1}\rangle=0, \quad \langle w_i^2 \rangle=1, \quad \langle w_i^4\rangle=\gamma^4, \quad \forall  i=1,\cdots , N, \forall k\in\mathbb{N}.
\end{equation}
This gives us the two-point function to be,
\begin{equation}\label{eq:ensemble-2pt}
    G^{(2)}_{\varphi\varphi}(X_1,X_2)= \frac{1}{N} \sum_{i,j=1}^N \langle w_i w_j \rangle \langle \Phi_i(X_1)\Phi_j(X_2)\rangle = \langle \Phi_i(X_1)\Phi_i(X_2)\rangle,
\end{equation}
with no sum on the $i$-index. So, starting with generic field of scaling dimension $\Delta_{\Phi}$, it gives us
\begin{equation}\label{eq:ensemble-2pt-generic}
     G^{(2)}_{\varphi\varphi}(x_1,x_2)= \frac{1}{x_{12}^{2\Delta_{\Phi}}}.
\end{equation}
Similarly, the four-point function is given by
\begin{equation}\label{eq:4pt-ensemble}
    G^{(4)}_{\varphi\varphi\varphi\varphi}(X_1,X_2,X_3,X_4)=\frac{\gamma^4}{N}G^{(4)}_{\Phi\Phi\Phi\Phi}(X_1,X_2,X_3,X_4)+\left(1-\frac{1}{N}\right)\left[G^{(2)}_{\Phi\Phi}(X_1,X_2)G^{(2)}_{\Phi\Phi}(X_3,X_4)+\text{perms}\right]
\end{equation}
which can be written down as,
\begin{equation}\label{eq:4pt-ensemble-2}
     G^{(4)}_{\varphi\varphi\varphi\varphi}(x_1,x_2,x_3,x_4)=(x_{12}^2x_{34}^2)^{-\Delta_{\Phi}} g_{\varphi}(u,v),
\end{equation}
where
\begin{equation}\label{eq:4pt-ensemble-guv}
    g_{\varphi}(u,v)=\frac{\gamma^4}{N}g_{\Phi}+\left(1-\frac{1}{N}\right) \left(1+u^{\Delta_{\Phi}}+\left(\frac{u}{v}\right)^{\Delta_{\Phi}}\right).
\end{equation}
We get a generalized free field of scaling dimension $\Delta_{\Phi}$ in the $N\rightarrow$ limit, including a free boson in $D=4$ setting $\Delta_{\Phi}=1$.

\section{Moments and Gram Matrices}\label{sec:moments}
We will now study the constraints on the moments of the parameter distribution $P(\Theta)$ imposed by the Gram matrices associated with our operators $\mathcal{O}_{m,n}$. Before we proceed, we will define some quantities for later convenience.

\begin{definition}[Normalized moments]\label{def:alpha}
We define the \emph{normalized moments} (ratio to Gaussian)
\begin{equation}
    \alpha_k := \frac{\mu_{2k}}{(2k-1)!!}\,,
\end{equation}
with the convention $\alpha_0 = 1$. The Gaussian distribution has $\alpha_k = 1$ for all $k$.
\end{definition}

\begin{definition}[Wick coefficient]
We define the \emph{Wick coefficient}
\begin{equation}\label{eq:wick_coeff}
    c^{(n)}_p(D) := n!\cdot R_p(D+2 + 2n)\,,
\end{equation}
so that the two-point function coefficient becomes
\begin{equation}
    G^{(n)}_{m,m'} = \alpha_k \cdot c^{(n)}_p(D)(X\cdot Y)^n\,,\qquad p = m+m',\quad k = p + n\,.
\end{equation}
\end{definition}

\begin{definition}[Gram matrix]
For fixed $n$ and operators $\{O_{0,n}, O_{1,n}, \ldots, O_{M,n}\}$, the Gram matrix is the $(M+1)\times(M+1)$ matrix with entries given by,
\begin{equation}
    \mathcal{G}^{(n)}_{m,m'} = \alpha_{m+m'+n}\cdot c^{(n)}_{m+m'}(D)\,.
\end{equation}
\end{definition}

\begin{proposition}[Hankel structure]
The Gram matrix $\mathcal{G}^{(n)}$ has Hankel structure, i.e., the entry $(m,m')$ depends only on $m+m'$:
\begin{equation}
    \mathcal{G}^{(n)}_{m,m'} = f_n(m+m')\,,\qquad f_n(p) = \alpha_{p+n}\cdot c^{(n)}_p(D)\,.
\end{equation}
\end{proposition}

\subsection{Warm-up with simple examples}
 For example, consider the operators $\mathcal{O}_{0,0}$, $\mathcal{O}_{1,0}$ and $\mathcal{O}_{2,0}$, where in $D=4$, we have
\begin{equation}\label{example:explicit-m012-n0}
    \begin{split}
        & G^{(2)}_{0,0;0,0}= 1 \\
        & G^{(2)}_{1,0;1,0}= \frac{\mu_4}{3}\left[ (D+2)^2+2(D+2) \right] = \frac{48}{3}\mu_4 \\
        & G^{(2)}_{2,0;2,0}= \frac{\mu_8}{105}\left[ (D+2)^4 + 12 (D+2)^3 + 44 (D+2)^2 + 48 (D+2) \right]= \frac{5760}{105} \mu_8 \\
        & G^{(2)}_{0,0;1,0} = \mu_2 (D+2) = 6 \mu_2 \\
        & G^{(2)}_{0,0;2,0} = \frac{\mu_4}{3} \left[ (D+2)^2 +2 (D+2) \right] = \frac{48}{3}\mu_4  \\
        & G^{(2)}_{1,0;2,0}= \frac{\mu_6}{15} \left[ (D+2)^3 +6 (D+2)^2 + 8(D+2) = \frac{480}{15} \mu_6 \right]
    \end{split}
\end{equation}

Let us now compute the determinants. For the $2\times2$ submatrices, the moments need to satisfy
\begin{equation}\label{eq:n0-2x2}
\begin{split}
    & \text{det} \mathcal{G}^{0}_{0,1} = 16\mu_4 - 36\mu_2^2 > 0, \quad \implies \mu_4 > \frac{9}{4} \mu_2^2, \\
    & \text{det} \mathcal{G}^{0}_{0,2}= \frac{5760}{105}\mu_8 - 256 \mu_4^2, \quad \implies \mu_8 > \frac{14}{3} \mu_4^2, \\
    & \text{det} \mathcal{G}^{0}_{1,2} = \frac{6144}{7} \mu_8 \mu_4 - 1024 \mu_6^2 > 0, \quad \implies \mu_8\mu_4> \frac{7}{6} \mu_6^2,
\end{split}
\end{equation}
and the $3\times3$ submatrix gives the condition,
\begin{equation}\label{eq:n0-3x3}
    \text{det} \mathcal{G}^{0}_{0,1,2} = -4096 \mu_{4}^{3} + 6144 \mu_{2} \mu_{4} \mu_{6} - 1024 \mu_{6}^{2} - \frac{13824}{7} \mu_{2}^{2} \mu_{8} + \frac{6144}{7} \mu_{4} \mu_{8} > 0
\end{equation}

In principle, there are infinitely many constraints on the moments that need to be satisfied. In the following subsection, we will give numerical evidence that the standard normal distribution satisfies all these constraints up to a high value of $m,n$ and use the numerical trends to extrapolate and conjecture the statistics at $m,n\rightarrow\infty$. We restrict ourselves to $D=4$.

\subsection{Results from numerical experiments}
Now we will present numerical and analytical evidence for the positive-definiteness of the Gram matrices. Recall that we restrict to $D+2 = 6$ throughout. As a starter, the Wick coefficients $c^{(n)}_p(D=4)$ for some values are given in table \ref{tab:wick-coeffs} below.
\begin{table}[h!]
\centering
\begin{tabular}{c|ccccc}
\hline
$p \backslash n$ & 0 & 1 & 2 & 3 & 4 \\ \hline
0 & 1 & 1 & 2 & 6 & 24 \\
1 & 6 & 8 & 20 & 72 & 336 \\
2 & 48 & 80 & 240 & 1008 & 5376 \\
3 & 480 & 960 & 3360 & 16128 & 96768 \\
4 & 5760 & 13440 & 53760 & 290304 & 1935360 \\ \hline
\end{tabular}
\caption{Wick coefficients $c^{(n)}_p(D=4)$ for some values of $p$ and $n$ in $D=4$.}
\label{tab:wick-coeffs}
\end{table}

There are additional properties of the Gram matrix and its submatrices that might be of interest. We have relegated all those details to appendix \ref{app:gram_matrix}. Returning to the question of positivity of Gram matrix, it can be shown that for the Gaussian distribution, the determinant is positive and growing (see table \ref{tab:Gram-determinants}).

\begin{table}[h!]
\centering
\begin{tabular}{c|rrrr}
\hline
Size & $n = 0$ & $n = 1$ & $n = 2$ & $n = 3$ \\ \hline
$1\times 1$ & 1 & 1 & 2 & 6 \\
$2\times 2$ & 12 & 16 & 80 & 864 \\
$3\times 3$ & $4{,}608$ & $10{,}240$ & $153{,}600$ & $7{,}927{,}296$ \\
$4\times 4$ & $5.31\times 10^7$ & $2.10\times 10^8$ & $1.03\times 10^{10}$ & $2.69\times 10^{12}$ \\
$5\times 5$ & $1.83\times 10^{13}$ & $1.37\times 10^{14}$ & $2.32\times 10^{16}$ & $3.30\times 10^{18}$ \\
$6\times 6$ & $1.90\times 10^{19}$ & $2.88\times 10^{20}$ & $1.76\times 10^{23}$ & $1.46\times 10^{25}$ \\ \hline
\end{tabular}
\caption{The determinants $\Delta^{(n)}_{k}$ of the $k\times k$ Gram matrices truncated at various $k$ for different $n$ for the Gaussian normal distribution.}
\label{tab:Gram-determinants}
\end{table}

We can also look at the normalized determinant (see table \ref{tab:gram_normal}), where
\begin{equation}
    \hat{\Delta}^{(n)}_{k} = \frac{\det\mathbf{G}^{(n)}_k} {\prod_{i=0}^{k-1} \mathbf{G}^{(n)}_{ii}}.
\end{equation}

\begin{table}[h!]
\centering
\begin{tabular}{c|cccc}
\hline
Size & $n=0$ & $n=1$ & $n=2$ & $n=3$ \\ \hline
$2\times 2$ & 0.2500 & 0.2000 & 0.1667 & 0.1429 \\
$3\times 3$ & 0.0139 & 0.0079 & 0.0050 & 0.0034 \\
$4\times 4$ & $1.05\times 10^{-4}$ & $3.67\times 10^{-5}$ & $1.45\times 10^{-5}$ & $6.34\times 10^{-6}$ \\
$5\times 5$ & $9.8\times 10^{-7}$ & $1.9\times 10^{-7}$ & $4.3\times 10^{-8}$ & $1.1\times 10^{-8}$ \\ \hline
\end{tabular}
\caption{The normalized determinant of the truncated $k\times k$ Gram matrix for various $k$ and $n$ for the Gaussian normal distribution.}
\label{tab:gram_normal}
\end{table}

The above analysis suggests that for the Gaussian normal distribution, the Gram matrix and the normalized Gram matrix are positive definite, satisfying our assertion \ref{ass:gram-sub-matrix}.

\section{Algebra from Correlation Functions}
\label{sec:algebra-correlators}
In this section, we will construct a $*$-algebra from the neural network conformal theory correlators. We will define two separate notions of inner product: the first is defined at finite separation on the parameter space. This inner product is defined for the well-definedness of the operators in the parameter space. In this definition, the space-time points are held fixed and conjugates do not act on them. We will define a second inner product where the conjugate operator acts on the space-time coordinates and this inner-product is a defined on the function space of fields. We will only formally define it and do not use it for any evaluation, but we will return to the non-positivity of this inner product later, to argue for the non-existence of Hilbert space for non-unitary theories.

Let us review some relevant concepts\footnote{For a self-contained review of algebraic quantum field theory, see \cite{Haag:1996hvx,Halvorson:2006wj}}. First we will define an inner product at finite separation. 

\begin{definition}[Inner-product on parameter space]
We have the bilinear product of $\mathcal{O}(X)$ and $\mathcal{O}'(Y)$ defined by the weighted overlap integral in the parameter space,
\begin{equation}\label{eq:overlap}
    || \mathcal{O}(X) \mathcal{O}'(Y) ||_p = \int d\theta P(\Theta) \mathcal{O}(X) \mathcal{O}'(Y),
\end{equation}
\end{definition}
where the subscript $p$ denotes that the bilinear product is defined on the parameter space and the finitely separated points $X$ and $Y$ are held fixed, and we can define a $\epsilon$-norm of the operator $\mathcal{O}(X)$ as \footnote{The definition of $\epsilon$-norm in eq. \ref{eq:norm} should not be confused with that of expectation value in eq. \ref{eq:exp}.}, 
\begin{equation}\label{eq:norm}
    || \mathcal{O}(X) ||_{\epsilon} = \sqrt{\int d\theta P(\Theta) \mathcal{O}(X)  \mathcal{O}(X+\epsilon)}.
\end{equation}

\begin{assertion}[Gram sub-matrix positivity]\label{ass:gram-sub-matrix}
The determinant of the Gram sub-matrix between any two operators is positive definite.
\end{assertion}

Clearly, our examples with the Gaussian normal distribution in section \ref{sec:moments} satisfy assertion \ref{ass:gram-sub-matrix} to any finite truncation of the operator spectrum at some $m_{max},n_{max}$. However, as $m_{max},n_{max}\rightarrow\infty$, the determinant of the Gram matrix approaches zero, and Gram matrix is not invertable anymore. This non-invertability is the source of the puzzle concerning mixed four-point functions. Nevertheless, with the Gaussian normal distribution, all our operators $\mathcal{O}_{m,n} $ are well-defined at any finite truncation. In the case of any other distribution, one has to remove the negative-normed and null states from the parameter-space.

Now we will discussion the Cauchy-Schwarz property of this inner product defined above. Consider two operators $\mathcal{O}_{m,n}(X)$ and $\mathcal{O}_{m',n'}(Y)$, and they necessarily satisfy,
    \begin{equation}
        G^{(2)}_{{m,n};{m',n'}}(X,Y) \leq \sqrt{G^{(2)}_{{m,n};{m,n}}(X,Y)} \sqrt{G^{(2)}_{{m',n'};{m',n'}}(X,Y)},
    \end{equation}
where we have used the symmetric property of the two-point function, i.e, 
\begin{equation}\label{prop:symm}
    G^{(2)}_{{m,n};{m',n'}}(X,Y)=G^{(2)}_{{m',n'};{m,n}}(X,Y).
\end{equation}

\begin{definition}[*-algebra]\label{def:star-algebra}
    Let $\mathcal{A}$ be an algebra. It is a *-algebra, if it has an anti-linear map $^* : \mathcal{A} \rightarrow \mathcal{A}$ taking $a\rightarrow a^*$, satisfying
    \begin{equation}
        (\mu a + \lambda b)^*= \bar{\mu} a^* + \bar{\lambda} b^*, \quad (a^*)^*=a, \quad (ab)^*=b^*a^*,
    \end{equation}
    for all $a,b\in\mathcal{A}$ and $\lambda,\mu\in\mathbb{C}$.
\end{definition}

In our case, in the parameter space with the space-time point $X$ held fixed, we have
\begin{equation}\label{eq:conjugate-X-fixed}
    \mathcal{O}^{*_p}(X) = \mathcal{O}(X), \quad \mathcal{O}(X)\mathcal{O}(Y)'=\mathcal{O}'(Y)\mathcal{O}(X),
\end{equation}
where the subscript $p$ in $*_p$ denotes that this involution is only the parameters, which trivially satisfies definition \ref{def:star-algebra}.

\begin{definition}[Unital *-algebra]\label{def:unital-star-algebra}
    Let $\mathcal{A}$ be a *-algebra. It is unital, if there exists a unique identity element $\mathbf{1}$, such that
    \begin{equation}
        \mathbf{1} \mathcal{O} = \mathcal{O}= \mathcal{O} \mathbf{1}.
    \end{equation}
\end{definition}

Clearly, we have $\mathcal{A}$ to be a unital *-algebra, and the operator $\mathcal{O}_{0,0}$ can be identified as $\mathbf{1}$.

\begin{definition}[C*-algebra]\label{def:C-star-algebra}
    A *-algebra $\mathcal{A}$ is a C*-algebra, if the norm satisfies
    \begin{equation}\label{eq:c-star-alg}
        ||a^*a||=||a||^2,
    \end{equation}
    for all $a\in\mathcal{A}$.
\end{definition}

Since we do not have bounded norms for every $a\in\mathcal{A}$, it is not a C*-algebra.

We will now define the weight, factor, trace and state as follows,
\begin{definition}[Linear functional]\label{def:linear-functional}
    We define a linear functional $\omega$ to be $\omega$: $\mathcal{A}\rightarrow\mathbb{C}$ satisfies,
    \begin{equation}
        \omega(\mu a + \lambda b) = \mu \omega(a) +\lambda \omega(b), \quad \omega(a^*)=(\omega(a))^*
    \end{equation}
    for all $\mu,\lambda \in \mathbb{C}$, $a,b\in\mathcal{A}$.   
\end{definition}

\begin{definition}[Weight]\label{def:weight} A weight is a linear functional $\omega$ that satisfies
\begin{equation}
    \omega(a^*a) \geq 0, \quad \forall a \in \mathcal{A}.
\end{equation}
\end{definition}

Note that a weight $\omega$ satisfies the Cauchy-Schwarz inequality,
\begin{equation}
    |\omega(a^*b)|^2 \leq |\omega(a^*a)||\omega(b^*b)|.
\end{equation}

We would like to include all operators $\mathcal{O}_{m,n}$ in our algebra $\mathcal{A}$. Therefore, we have to ensure that they are all linearly independent,i.e.,
\begin{equation}
    || \sum_m c_m \mathcal{O}_{m,n} ||^2 = 0, \quad \text{iff}\quad c_m=0 \quad \forall m.
\end{equation}
For simplicity, let us consider two operators $\mathcal{O}_{m,n}$ and $\mathcal{O}_{m',n}$. WLOG, let's consider,
\begin{equation}
    || \mathcal{O}_{m',n} - c^{(n)}_{m,m'} \mathcal{O}_{m,n} ||^2 > 0.
\end{equation}
This leaves us with the condition,
\begin{equation}\label{eq:2op-condition}
    || \mathcal{O}_{m',n} \mathcal{O}_{m',n} || + Re(c^{(n)}_{m,m'})^2 || \mathcal{O}_{m,n} \mathcal{O}_{m,n} ||- 2 Re(c^{(n)}_{m,m'}) ||\mathcal{O}_{m,n} \mathcal{O}_{m',n}|| > 0,
\end{equation}
which is automatically satisfied by the assertion \ref{ass:gram-sub-matrix}. To be more explicit, let us solve for the roots of the quadratic equation in $Re(c^{(n)}_{m,m'})$ saturating \ref{eq:2op-condition},
\begin{equation}\label{eq:2gram-condition}
\begin{aligned}
    c^{(n)\pm}_{m,m'}&= \frac{2G^{(2)}_{m,n;m'n} \pm \sqrt{4 (G^{(2)}_{m,n;m'n})^2 - 4 G^{(2)}_{m,n;m,n} G^{(2)}_{m',n;m',n}}}{2 G^{(2)}_{m',n;m',n}} \\
    & = G^{(2)}_{m,n;m'n} \pm i \sqrt{\text{det} \mathcal{G}^{(n)}_{m,m'}},
\end{aligned}
\end{equation}
where $\mathcal{G}^{(n)}_{m,m'}$ is the Gram (sub-)matrix for operators $\mathcal{O}_{m,n}$ and $\mathcal{O}_{m',n}$. As long as det$\mathcal{G}^{(n)}_{m,m'}>0$, there is no solution for equation \ref{eq:2op-condition}.

More broadly, we require the complete Gram matrix (and any $k\times k$ sub-matrix) to have positive definite determinant. Nonetheless, this exercise leaves us with an infinite dimensional *-algebra over the complex field $\mathbb{C}$ generated by operators $\mathcal{O}_{m,n}$.

It is important to note that the positivity of Gram matrix above is defined with respect to the parameter space, and not the space-time. Therefore, this positivity of Gram matrix holds for some arbitrary but fixed finite separation. However, from an axiomatic quantum field theory point of view, we are more interested in the positivity in space-time.

Now we will define the inner product on function space which is the inner-product of fields \footnote{In the light of parameter space-function space duality \cite{Maiti:2021fpy}, we have tweaked the definition of parameter space inner-product, and therefore, the duality does not hold here.}, and is not necessarily positive definite, as our non-unitary theories can contain negative-normed states.
\begin{definition}[Inner product on function space]
Given two fields $\mathcal{O}_1(X)$ and $\mathcal{O}_2(Y)$, the inner-product between them is defined as,
\begin{equation}
    \langle \mathcal{O}_1(X), \mathcal{O}_2(Y) \rangle = \int d\mathcal{O} \rho({\mathcal{O}}) \mathcal{} \mathcal{O}_1(X)^{\dagger} \mathcal{O}_2(Y),
\end{equation}  
where the measure $d\mathcal{O}$ is ill-defined and we do not know the explicit form of the density $\rho(\mathcal{O})$, and therefore, one may rewrite this integral in terms of the parameter space variables \cite{Maiti:2021fpy}, which is an well-defined integral,
\begin{equation}
   \langle \mathcal{O}_1(X), \mathcal{O}_2(Y) \rangle =  \int d\theta P(\Theta) \mathcal{} \mathcal{O}_1(X)^{\dagger} \mathcal{O}_2(Y).
\end{equation}
\end{definition}

Note that now, we can have, more generally,
\begin{equation}
    \mathcal{O}^{\dagger}(X)\neq \mathcal{O}(X),
\end{equation}
unlike \ref{eq:conjugate-X-fixed}. This is because in an Euclidean CFT, time is imaginary and conjugate acts like a reflection. We will see that for non-unitary theories, this inner-product (including, the norm) is not positive definite. However, that should not deter us from constructing the algebra from correlators, as we will see in the next section. We will refer to this complex conjugation $^\dagger$ as the involution $^*$ for our algebra, i.e., which satisfies the definition of a $*$-algebra in \ref{def:star-algebra}.

\subsection{Axiomatic Quantum Field Theories without Reflection Positivity}
In this subsection, we will review the Osterwalder–Schrader (OS) axioms \cite{Osterwalder:1974tc,Glimm:1987ng} for Euclidean quantum field theories and then show that reflection positivity is not guaranteed in our theory. It is well-known that Euclidean conformal theories with negative scaling dimensions violating the unitarity lower bound of operator dimensions are non-unitary theories, and do not satisfy reflection positivity \cite{Mack:1975je}. These theories contain negative-norm states, and strictly speaking, one cannot construct Hilbert spaces, which by definition require the norm to be positive. However, the work of \cite{Jakobczyk:1984ip,Jakobczyk:1987zw,Jakobczyk:1987zx} gives us a technique to handle some of these theories. We make an important clarification here that the neural network theories are in general non-local. As argued in the appendix of \cite{Halverson:2024axc}, that for specific choice of moments, we can get local correlators. In the following discussion, we will be considering our choice of moments satisfy the conditions for local correlation functions. Moreover, \cite{Halverson:2024axc} introduced generalized free fields in large-$N$ limit of an ensemble of neural networks which we reviewd in \ref{subsec:larg-N}. The following discussion also applies to such generalized free fields which do not have a local stress tensor but satisfy the axioms. \footnote{Nevertheless, the purpose of this section is to reminder the reader on the existence of state-space for non-unitary theories and we will not adopt the language of Haag-Kastler and will not construct \emph{local} nets for spacetime diamonds. Instead we will use the conformal group and state-operator correspondence to define our algebra.}

Consider $\{S_n\}$ to be the set of Schwinger functions (whose analytic continuation are the Wightman functions $\{W_n\}$ \cite{osti_4606723}), satisfying the following axioms in $D=4$ dimensions:

\begin{enumerate}[label=OS\arabic*]
      \item \emph{Temperedness}: For each $n$, $S_n$ is a tempered distribution on the space of test functions vanishing on diagonals,
    \[
        S_n \in \mathcal{S}_0'(\mathbb{R}^{4n}),
    \]
    and satisfies the Hermiticity condition,
    \[
        \overline{S_n(f)} = S_n(\mathbb{T} f^*),
    \]
    where $\mathbb{T} f(x_1,\ldots,x_n) = f(rx_1,\ldots,rx_n)$, $\;r(x^0,\mathbf{x}) = (-x^0,\mathbf{x})$ is Euclidean time reflection, and $f^*(x_1,\ldots,x_n) = \overline{f(x_n,\ldots,x_1)}$.
    
    \item \emph{Euclidean covariance}: For every $(a,R)\in \mathbb{R}^4 \rtimes SO(4)$, \[S_n(x_1,\dots,x_n)=S_n(Rx_1 + a,\dots, Rx_n + a).\]
    
    \item \emph{Reflection positivity}: Let $\mathcal{B}_+$ be the Borchers algebra over $\mathcal{S}_+(\mathbb{R}^{4n})$, the space of test functions with strictly ordered Euclidean times $0 < x_1^0 < x_2^0 < \cdots < x_n^0$. Then for all $F_+ \in \mathcal{B}_+$,
    \[
        S(\mathbb{T} F_+^* \times F_+) \;\equiv\; \sum_{n,m} S_{n+m}(\mathbb{T} f_n^* \times f_m) \;\geq\; 0.
    \]
    
    \item \emph{Symmetry}: For any permutation $\pi \in S_n$, \[S_n(x_{\pi(1)},\dots,x_{\pi(n)})=S_n(x_1,\dots,x_n).\]
\end{enumerate}

It is clear that for non-unitary quantum field theories, $OS3$ does not hold and must be dropped. It is still possible to axiomatically define such a QFT with some modifications \cite{Jakobczyk:1984ip,Jakobczyk:1987zw,Jakobczyk:1987zx}. It will be useful for our purpose to replace $OS3$ with
\begin{enumerate}[start=3,label=OS\arabic*']
    \item \emph{Krein positivity}: There exists a mapping $\alpha_s \colon \mathcal{B}_+ \to \mathcal{B}_+$ such that for all $F_+, G_+ \in \mathcal{B}_+$:
    \begin{enumerate}[label=(\roman*)]
        \item $S\bigl(\{\mathbb{T}\,\alpha_s(\alpha_s(F_+))\}^* \times G_+\bigr) = S\bigl(\{\mathbb{T} F_+\}^* \times G_+\bigr)$;
        \item $S\bigl(\{\mathbb{T}\,\alpha_s(F_+)\}^* \times F_+\bigr) \geq 0$;
        \item $S\bigl(\{\mathbb{T}\,\alpha_s(F_+)\}^* \times G_+\bigr) = S\bigl(\{\mathbb{T} F_+\}^* \times \alpha_s(G_+)\bigr)$;
        \item $p_{\alpha_s}(F_+) \equiv S\bigl(\{\mathbb{T}\,\alpha_s(F_+)\}^* \times F_+\bigr)^{1/2}$ is continuous in the $\mathcal{B}_+$-topology,
    \end{enumerate}
    where the seminorm $p_{\alpha_s}$ is non-degenerate, $\ker p_{\alpha_s} = \mathcal{N}_S$, where $\mathcal{N}_S$ is the Schwinger kernel. The completion of the Euclidean state space $\mathcal{D}^S = \mathcal{B}_+ / \mathcal{N}_S$ in the Hilbert topology of $p_{\alpha_s}$ is a Krein space $\mathcal{K}^S$ with metric operator $\eta_S$ defined by $\eta_S [F_+]_S = [\alpha_s(F_+)]_S$, satisfying $\eta_S^2 = \mathbf{1}$.
\end{enumerate}

We do not prove that our theory satisfies $OS3'$, for now we will assume that it does and will remind the reader whenever this assumption is used. The point of the above discussion was to shed light on the fact that we have a state-space and (conjecturally) it has the structure of a Krein-space. Now we are ready to discuss the algebra and construct the state-space from correlators.

\subsection{The Algebra and Construction of State Space}
We will first discuss the algebra of fields which is different from the algebra of correlators in our theory. In other words, the algebras before and after averaging have distinct characteristics. This distinction manifests as the fields by themselves upon multiplication follow the \emph{word} algebra, whereas after taking expectation values demonstrate a richer structure.

\subsubsection{Algebra of fields}
Consider two local operators $\mathcal{O}_{m_1,n_1}(X)$ and $\mathcal{O}_{m_2,n_2}(Y)$, and define scalar multiplication as a product (denoted by $\odot$).

For two local operators separated by finite distance, we get,
\begin{equation}
    \label{eq:field_algebra_bilocal}
    \begin{aligned}
     \mathcal{O}_{m_1,n_1}(X) \odot \mathcal{O}_{m_2,n_2}(Y) & = (\theta\cdot\theta)^{m_1}(\theta \cdot X)^{n_1} (\theta\cdot\theta)^{m_2}(\theta \cdot Y^{n_2} \\
     & =  (\theta\cdot\theta)^{m_1+m_2} (\theta\cdot X)^{n_1} (\theta\cdot Y)^{n_2} \\
    \end{aligned}
\end{equation},
where the RHS is a bilocal operator, and a repeated multiplication will simply generating multilocal operators. However, we are more interested in local operators. In particular, we have the following local operator the limit $Y\rightarrow X$,
\begin{equation}
    \label{eq:field_algebra_local}
    \begin{aligned}
     \lim_{\varepsilon\rightarrow 0}   \mathcal{O}_{m_1,n_1}(X) \odot \mathcal{O}_{m_2,n_2}(X+\varepsilon) & = (\theta\cdot\theta)^{m_1}(\theta \cdot X)^{n_1} (\theta\cdot\theta)^{m_2}(\theta \cdot (X+\varepsilon))^{n_2} \\
     & \simeq  (\theta\cdot\theta)^{m_1+m_2} (\theta\cdot X)^{n_1+n_2} \\
    & = \mathcal{O}_{m_1+m_2,n_1+n_2}(X)
    \end{aligned}
\end{equation}
where in the second equality, we have suppressed terms in powers of $\epsilon$. More generally, for a \emph{finite} number of operators $\mathcal{O}_{m_i,n_i}$(X), we have
\begin{equation}\label{eq:field_algebra_multi}
    \prod_{i;\odot}  \mathcal{O}_{m_i,n_i}(X) = \mathcal{O}_{M,N}(X),
\end{equation}
where $M=\sum_i m_i$ and $N=\sum_i n_i$. It is trivial to check that this is closed under multiplication. This algebra is commutative and associative.

\subsubsection{Algebra of correlators}

The correlation functions constructed by this neural network approach obey conformal symmetry. It is then natural to ask if these correlation functions can be thought of as vacuum expectation values of some operator algebra. We now argue that indeed these correlation functions can be thought of as an algebraic state on an abstract $*$-algebra.

Since we are dealing with a {\it conformal} (field) theory, rather than an arbitrary quantum field theory, this algebra can be described fairly explicitly in terms of the generators of conformal symmetries and a set of primary operators. 

Conventionally in AQFT, the fundamental object of study is the algebra of (the causal completion of) ball-shaped subregions, and one views algebras associated to general subregions as built from these fundamental building blocks. In the case of a conformal theory, the existence of dilatation symmetry greatly simplifies the analysis.

Let us denote the algebra we would like to describe by $\mathcal{A}_{\rm CFT}$. Rather than describing $\mathcal{A}_{\rm CFT}$ using local operators at arbitrary spacetime points $\mathcal{O}(x)$, we use a set of primary operators, $\mathcal{P}_0$, located at the origin (here we have seen $\mathcal{P}_0 = \{\mathcal{O}_{0,n}, ~\forall n \in \mathbb{N}\}$), descendants ($\mathcal{D}_0 = \{\partial_{i_1} \cdot\cdot\cdot \partial_{i_n} \mathcal{O}_{\Delta}(0)\}$), and the conformal algebra generated by $\mathcal{G}_{\rm conf.} = \{P_i, K_i, M_{ij}, D\}$.
We define a vector space $\mathcal{V}$ as the set of all finite linear combinations (with complex coefficients) of finite products of elements of $\mathcal{P}_0$, $\mathcal{D}_0$, and $\mathcal{G}_{\rm conf.}$. 

To obtain an algebra, we must add algebraic relations  between these vector space elements. These relations are described by the usual relations between the conformal generators, the OPE between primary operators, and additionally
\begin{equation}
    [P_i, \mathcal{O}_{\Delta}] = \partial_i \mathcal{O}_{\Delta}, [K_i, \mathcal{O}_{\Delta}] = 0, [M_{ij}, \mathcal{O}_{\Delta}] = 0, [D, \mathcal{O}_{\Delta}] = \Delta \mathcal{O}_{\Delta} \ ,
\end{equation}
where here $\Delta \in -\mathbb{N}$.

We are interested in describing a state space in which these operators act and we will see that this requires the definition of an involution on the algebra which will become analogous to the adjoint operation when the operators are represented on a state space. The state space we obtain depends on the quantization scheme (radial versus flat) and correspondingly so does the defintion of the involution.
We focus on radial quantization below.

We have not yet described a $*$-algebra as we have not defined any involution on the operators we are discussing. To define the involution in a manner consistent with radial quantization, we must introduce a new set of operators, $\mathcal{P}_{\infty}$ (here $\mathcal{P}_{\infty} = \{\mathcal{O}_{0,n}^*, ~n \in \mathbb{N} \}$) simply defined to be the images of $\mathcal{P}_0$ under the involution. In conventional language, these can be thought of as primaries at infinity.\footnote{We can think of $\mathcal{O}_{0,n}^*$ as $\lim_{y \to \infty} y^{-2\Delta} \mathcal{O}_{0,n}(y)$. 
} Additionally, on the conformal generators we have
\begin{equation}
    P_i^* = K_i,~ M_{ij}^* = -M_{ij},~ D^* = D \ .
\end{equation}
The involution is extended to the entire algebra by conjugate linearity and through the relation $(xy)^* = y^* x^*,~\forall x,y \in \mathcal{A}_{\rm CFT}$. 

Now that $\mathcal{A}_{\rm CFT}$ is a genuine $*$-algebra, we view the correlation functions of the neural network conformal theory as values taken on by a linear functional $\omega: \mathcal{A}_{\rm CFT} \to \mathbb{C}$. The state space is constructed using a generalization of the GNS construction \cite{Hofmann:1993dm,Hofmann:1995ue}. The reason why we must use a generalized GNS construction is related to the ``non-unitarity'' of this theory. We now recall a well-known argument establishing that negative scaling dimensions give rise to negative norm states.

Consider the operator $X = P_0 \mathcal{O}_{\Delta} \in \mathcal{A}_{\rm CFT}$. In the usual GNS construction we would have a state $\ket{X}$ with norm squared given by $||\ket{X}||^2 = \omega(X^*X)$. Using the algebraic relations of $\mathcal{A}_{\rm CFT}$ one can show that 
\begin{equation}
    X^*X = 2\Delta \mathcal{O}_{\Delta}^* \mathcal{O}_{\Delta} + 2 \mathcal{O}_{\Delta}^* \mathcal{O}_{\Delta} D + \mathcal{O}_{\Delta}^* P_0 \mathcal{O}_{\Delta} K_0.
\end{equation}
To interpret the functional $\omega$ as the conformal vaccum state, it must vanish whenever the rightmost operator is $K_i,~ M_{ij},$ or $D$ or whenever the leftmost operator is $P_i,~ M_{ij},$ or $D$. Thus, one obtains
\begin{equation}
    \omega(X^*X) = 2\Delta \omega(\mathcal{O}_{\Delta}^* \mathcal{O}_{\Delta}) = \lim_{y\to \infty} y^{2\Delta} \langle \mathcal{O}_{\Delta}(y) \mathcal{O}_{\Delta}(0) \rangle = 2\Delta.
\end{equation}
Thus the state $\ket{X}$ that we would like to define in the usual GNS procedure has negative norm. In AQFT language, the linear functional $\omega$ which is described by correlation functions is not a weight,\footnote{Weights $\rho: \mathcal{A}_{CFT} \to \mathbb{C}$ must satisfy $\rho(Y^*Y) \geq 0,~ \forall Y \in \mathcal{A}_{\rm CFT}$.} thus the standard GNS construction does not apply.

Instead a weaker representation theorem due to Scheibe applies and establishes the existence of a vector space on which the operators can be represented, but no additional structure beyond that of a vector space is guaranteed.

Since the standard GNS construction does not apply, there is not a unique $*$-representation of this algebra on a Hilbert space. However, if we assume that the correlation functions obey a certain ``partial majorant'' condition, then there is a ``$J$-representation'' of the algebra on a Krein space.

\section{Discussions}
\label{sec:discuss}
\subsection{Lessons for conformal theories}
The construction of Section \ref{sec:algebra-correlators} describes the algebra $\mathcal{A}_{\rm CFT}$ not through local operators $\mathcal{O}(x)$ at arbitrary spacetime points, but through a small set of data attached to a single point, i.e., the root primaries $\mathcal{P}_0$, their images at infinity $\mathcal{P}_\infty$, the descendants $\mathcal{D}_0$, and the conformal generators $\mathcal{G}_{\rm conf.}$. It is worth emphasizing that this is not merely a convenient bookkeeping device but reflects a structural feature that distinguishes conformal theories from generic quantum field theories.

In the algebraic (Haag--Kastler) formulation of a generic QFT
\cite{Haag:1996hvx,Halvorson:2006wj}, the primitive datum is a \emph{net} of
operator algebras $O \mapsto \mathcal{A}(O)$ assigning to each causal diamond
(double cone) $O$ the algebra of observables localized in it. The global
(quasilocal) algebra is then assembled \emph{from} the regional ones, i.e., it is the $C^*$-algebra generated by $\bigvee_O \mathcal{A}(O)$ --- subject to isotony, microcausality, and covariance. The content of the theory lives in the relations between regions, where the local algebras are type III$_1$ factors \cite{Buchholz:2019rem,Brunetti:2021wev,Fredenhagen:2015utr}, there is no vacuum vector localized in any proper subregion (Reeh-Schlieder), and the vacuum modular flow of a region is the engine of the dynamics. Crucially, knowing one $\mathcal{A}(O)$ does not hand you the others; one genuinely needs the whole net.

Conformal symmetry collapses this picture. The conformal group acts transitively on the set of causal diamonds, so all the regional algebras $\mathcal{A}(O)$ are conformally conjugate and the net is homogeneous. The decisive extra ingredient is the dilatation $D$, where the unitary $e^{is D}$ maps the algebra of a diamond of one size to that of a diamond of a different size about the same point. There is no intrinsic scale separating ``large'' from ``small'' regions, so the algebra of all of space is conjugate to the algebra of an arbitrarily small neighbourhood of a point. This is the precise sense in which scale invariance lets one ``shrink the global down to the local'', so rather than specifying a spacetime net, it suffices to specify the operator data at a single point and let the conformal group transport it. Radial quantization makes this operational in the sense that the dilatation operator is the radial Hamiltonian and the state--operator map packages the local data at the origin into states, and it is exactly this data that $\mathcal{A}_{\rm CFT}$ encodes. The net
$O \mapsto \mathcal{A}(O)$ is then a \emph{derived} object, reconstructed by acting with (exponentials of) the conformal generators on the point data, rather than a fundamental input. For the double cone this geometric organization is also visible at the level of modular theory, where the vacuum modular flow acts as a conformal flow \cite{Hislop:1981uh,Hislop:1988yg}.

\subsection{Lessons for non-unitary theories}
The Haag--Kastler/OS framework presupposes positivity. Reflection positivity (OS3) is what manufactures a Hilbert space from the Euclidean correlators, and only on that Hilbert space do the regional algebras become $C^*$- or von Neumann algebras of bounded operators; the entire ``net of type III factors'' picture rests on this input. Our theory removes that input. The scaling dimensions $\Delta = -n \le 0$ violate the unitarity bound \cite{Mack:1975je}, and as argued in Section \ref{sec:algebra-correlators} the correlator functional $\omega$ is \emph{not} a weight and can exhibit a negative-norm state. Several pillars of the standard construction therefore fall away. The GNS construction does not apply, so there is no canonical Hilbert space and no canonical representation. What remains is only Scheibe's weaker theorem, guaranteeing a representation on a bare vector space. The local algebras are not $C^*$-algebras to begin with, since the
norm \eqref{eq:norm} is unbounded, so the von Neumann completion is simply unavailable.

We are left with the algebraic skeleton of the $*$-algebra $\mathcal{A}_{\rm CFT}$ consisting of generators, relations and involution. The correlators realize a (sign-indefinite) linear functional on it. The conformal
organization discussed above, including the reduction to point data via dilatation, never used positivity and so carries over verbatim: the recipe for constructing the algebra from correlators is robust against non-unitarity. It is only the \emph{representation}, i.e., the passage from the abstract algebra to operators on an inner-product space, that degrades. The natural replacement for the Hilbert space is an indefinite-metric Krein space: OS3 is traded for Krein positivity (OS3$'$), and under the assumed partial-majorant condition one obtains a $J$-representation on a Krein space $\mathcal{K}^S$ with metric operator $\eta_S$, $\eta_S^2 = \mathbf{1}$
\cite{Jakobczyk:1984ip,Jakobczyk:1987zw,Jakobczyk:1987zx}. The indefinite metric is the controlled price of the negative-norm states, i.e., in place of a positive inner product, one keeps a decomposition into positive and negative subspaces together with a well-defined $\eta_S$.

\subsection{Lessons for neural-network conformal theories}
A point that deserves emphasis in our neural-network construction of non-unitary conformal theories (since it is easy to conflate) is that two distinct notions of positivity appear in this paper. The Gram-matrix positivity of Sections 3 and 4 is positivity in \emph{parameter space} --- positivity of the $P(\Theta)$-weighted overlaps at fixed spacetime separation, and it is what makes the operators $\mathcal{O}_{m,n}$ linearly independent and endows the $*$-algebra with a non-degenerate structure. It is \emph{not} spacetime reflection positivity. A positive-definite parameter-space Gram matrix is entirely compatible with a non-positive $\omega$ on the spacetime algebra; the former is a statement about the distribution of neurons, the latter about unitarity of the resulting field theory. Establishing that the Gram matrices are positive-definite (as we do for the Gaussian case) therefore certifies that the algebra is well-posed, but says nothing about unitarity, which indeed fails.

This leaves several questions open. Whether the correlators actually satisfy OS3$'$ and the partial-majorant condition is assumed rather than proved here; locality of the correlators (available only for special moments, as noted in the appendix of \cite{Halverson:2024axc}) is a prerequisite, and the existence of the Krein structure beyond that is not settled. Even granting a $J$-representation, the appropriate ``type'' classification of these indefinite-metric algebras, i.e., a Krein-space analogue of the type III story, is unclear and unsure if such a structure even exists. The cleanest setting in which to make all of this explicit is the large-$N$ generalized free field of Section \ref{subsec:larg-N}, starting from a generalized free field with $\Delta = -n < 0$, which is a quasi-free non-unitary theory, and its Krein space, metric operator, and $J$-representation should be constructible in closed form, providing a concrete testbed for the structures conjectured above.

\section*{Acknowledgements}
We thank Jim Halverson, Sridip Pal, Sylvain Ribault and David Simmons-Duffin for discussions. We thank Anthropic's Claude for assisting with the codes used for the numerical calculations. This research was supported in part by grant NSF PHY-2309135 to the Kavli Institute for Theoretical Physics (KITP) and Cooperative Agreement PHY-2019786 (The NSF AI Institute for Artificial Intelligence and Fundamental Interactions. J.N. would like to thank the KITP for hospitality and accommodation during the Graduate Fellowship program.

\appendix

\section*{Two-point function derivation via Wick contractions}
\label{app:a}
The two-point function involves the expectation
\begin{equation}
G^{(2)}_{m,n;m',n}(X,Y) = \langle O_{m,n}(X) \, O_{m',n}(Y) \rangle = \langle (\theta \cdot \theta)^{m+m'} (\theta \cdot X)^n (\theta \cdot Y)^n \rangle.
\end{equation}

This contains $2(m + m') + 2n = 2k$ factors of $\theta$, which must be Wick contracted pairwise upto null cone constraints that we discuss below.

\subsubsection*{Null Cone Constraint}
On the null cone, $X \cdot X = Y \cdot Y = 0$. This eliminates Wick contractions that would produce:
\begin{itemize}
\item $X \cdot X$ terms (from pairing two $\theta$s contracted with $X$)
\item $Y \cdot Y$ terms (from pairing two $\theta$s contracted with $Y$)
\end{itemize}

\subsubsection*{Surviving Contractions}
The only surviving contractions have the following structure:
\begin{enumerate}
\item Each of the $n$ indices from $(\theta \cdot X)^n$ pairs with exactly one index from $(\theta \cdot Y)^n$, contributing $(X \cdot Y)^n$. There are $n!$ ways to form these pairings.

\item The remaining $2p$ indices from $(\theta \cdot \theta)^p$ contract among themselves.
\end{enumerate}

\subsubsection*{Dimension Shift}
When the $(\theta \cdot \theta)^p$ indices contract in the presence of $2n$ indices already committed to $X$-$Y$ pairings, the effective dimension shifts from $d$ to $d + 2n$ where $d=D+2$. This gives:
\begin{equation}
\langle (\theta \cdot \theta)^p \rangle_{\text{eff}} = R_p(d + 2n).
\end{equation}

\subsubsection*{Full Result}
Combining all factors:
\begin{equation}
G^{(2)}_{m,n;m',n} = \frac{\mu_{2k}}{(2k-1)!!} \cdot \underbrace{n!}_{\text{$X$-$Y$ pairings}} \cdot \underbrace{R_p(d+2n)}_{\text{$\theta$-$\theta$ contractions}} \cdot (X \cdot Y)^n.
\end{equation}

\subsubsection*{Special Case: $n = 0$}

For $n = 0$, the formula reduces to
\begin{equation}
G^{(2)}_{m,0;m',0} = \frac{\mu_{2p}}{(2p-1)!!} \cdot R_p(d).
\end{equation}

\section*{Some properties of the Gram Matrix}
\label{app:gram_matrix}
There are additional structure within different submatrices. For example, the $2\times 2$ Gram matrix for $m = 0, 1$ at fixed $n$ is
\begin{equation}
    \mathbf{G}^{(n)}_{2\times 2} = \begin{pmatrix} \alpha_n \cdot c^{(n)}_0 & \alpha_{n+1}\cdot c^{(n)}_1 \\ \alpha_{n+1}\cdot c^{(n)}_1 & \alpha_{n+2}\cdot c^{(n)}_2 \end{pmatrix}\,.
\end{equation}
Positive-definiteness of this matrix requires
\begin{equation}\label{eq:2x2_constraint}
    \alpha_n\,\alpha_{n+2} > \frac{[c^{(n)}_1]^2}{c^{(n)}_0\, c^{(n)}_2}\;\alpha_{n+1}^2 = \frac{D+2 + 2n}{D+4 + 2n}\;\alpha_{n+1}^2\,.
\end{equation}

\begin{proposition}[Chain of $2\times 2$ constraints]
Defining $\rho_n := (D+2+2n)/(D+2+2n+2)$, the $2\times 2$ positivity conditions form the chain
\begin{equation}\label{eq:chain}
    \alpha_n\,\alpha_{n+2} > \rho_n\,\alpha_{n+1}^2\,,\qquad n = 0, 1, 2, \ldots
\end{equation}
with $\rho_n \nearrow 1$ as $n \to \infty$. Explicitly at $D+2 = 6$:
\begin{table}[h!]
\centering
\begin{tabular}{c|cc}
\hline
$n$ & $\rho_n$ & Constraint \\ \hline
0 & $3/4$ & $\alpha_0\,\alpha_2 > \tfrac{3}{4}\,\alpha_1^2$ \\
1 & $4/5$ & $\alpha_1\,\alpha_3 > \tfrac{4}{5}\,\alpha_2^2$ \\
2 & $5/6$ & $\alpha_2\,\alpha_4 > \tfrac{5}{6}\,\alpha_3^2$ \\
$n$ & $\frac{n+3}{n+4}$ & $\alpha_n\,\alpha_{n+2} > \frac{n+3}{n+4}\,\alpha_{n+1}^2$ \\ \hline
\end{tabular}
\end{table}
As $\rho_n \to 1$, these approach strict log-convexity of the sequence $\{\alpha_k\}$.
\end{proposition}

\bibliographystyle{JHEP}
\bibliography{main}

@article{Halverson:2024axc,
    author = "Halverson, James and Naskar, Joydeep and Tian, Jiahua",
    title = "{Conformal fields from neural networks}",
    eprint = "2409.12222",
    archivePrefix = "arXiv",
    primaryClass = "hep-th",
    doi = "10.1007/JHEP10(2025)039",
    journal = "JHEP",
    volume = "10",
    pages = "039",
    year = "2025"
}

@article{Halverson:2020trp,
    author = "Halverson, James and Maiti, Anindita and Stoner, Keegan",
    title = "{Neural Networks and Quantum Field Theory}",
    eprint = "2008.08601",
    archivePrefix = "arXiv",
    primaryClass = "cs.LG",
    doi = "10.1088/2632-2153/abeca3",
    journal = "Mach. Learn. Sci. Tech.",
    volume = "2",
    number = "3",
    pages = "035002",
    year = "2021"
}

@article{Halverson:2021aot,
    author = "Halverson, James",
    title = "{Building Quantum Field Theories Out of Neurons}",
    eprint = "2112.04527",
    archivePrefix = "arXiv",
    primaryClass = "hep-th",
    month = "12",
    year = "2021"
}

@article{Ferko:2026axm,
    author = "Ferko, Christian and Halverson, James and Mutchler, Aaron",
    title = "{Universality of Neural Network Field Theory}",
    eprint = "2601.14453",
    archivePrefix = "arXiv",
    primaryClass = "hep-th",
    month = "1",
    year = "2026"
}

@article{Frank:2026bui,
    author = "Frank, Samuel and Halverson, James",
    title = "{String Theory from Infinite Width Neural Networks}",
    eprint = "2601.06249",
    archivePrefix = "arXiv",
    primaryClass = "hep-th",
    month = "1",
    year = "2026"
}

@article{Frank:2025zuk,
    author = "Frank, Samuel and Halverson, James and Maiti, Anindita and Ruehle, Fabian",
    title = "{Fermions and Supersymmetry in Neural Network Field Theories}",
    eprint = "2511.16741",
    archivePrefix = "arXiv",
    primaryClass = "hep-th",
    month = "11",
    year = "2025"
}

@article{Ferko:2025ogz,
    author = "Ferko, Christian and Halverson, James",
    title = "{Quantum mechanics and neural networks}",
    eprint = "2504.05462",
    archivePrefix = "arXiv",
    primaryClass = "hep-th",
    doi = "10.1088/2632-2153/ae3105",
    journal = "Mach. Learn. Sci. Tech.",
    volume = "7",
    number = "1",
    pages = "015002",
    year = "2026"
}

@article{Demirtas:2023fir,
    author = "Demirtas, Mehmet and Halverson, James and Maiti, Anindita and Schwartz, Matthew D. and Stoner, Keegan",
    title = "{Neural network field theories: non-Gaussianity, actions, and locality}",
    eprint = "2307.03223",
    archivePrefix = "arXiv",
    primaryClass = "hep-th",
    doi = "10.1088/2632-2153/ad17d3",
    journal = "Mach. Learn. Sci. Tech.",
    volume = "5",
    number = "1",
    pages = "015002",
    year = "2024"
}

@article{Maiti:2021fpy,
    author = "Maiti, Anindita and Stoner, Keegan and Halverson, James",
    title = "{Symmetry-via-Duality: Invariant Neural Network Densities from Parameter-Space Correlators}",
    eprint = "2106.00694",
    archivePrefix = "arXiv",
    primaryClass = "cs.LG",
    month = "6",
    year = "2021"
}

@article{Halverson:2024hax,
    author = "Halverson, Jim",
    title = "{TASI Lectures on Physics for Machine Learning}",
    eprint = "2408.00082",
    archivePrefix = "arXiv",
    primaryClass = "hep-th",
    month = "7",
    year = "2024"
}

@article{Ageev:2026ofv,
    author = "Ageev, Dmitry S. and Ageeva, Yulia A.",
    title = "{Excited String States and D-branes from Infinite Width Neural Networks}",
    eprint = "2602.10214",
    archivePrefix = "arXiv",
    primaryClass = "hep-th",
    month = "2",
    year = "2026"
}

@article{Robinson:2025ybg,
    author = "Robinson, Brandon",
    title = "{Virasoro Symmetry in Neural Network Field Theories}",
    eprint = "2512.24420",
    archivePrefix = "arXiv",
    primaryClass = "hep-th",
    month = "12",
    year = "2025"
}

@article{Capuozzo:2025ozt,
    author = "Capuozzo, Pietro and Robinson, Brandon and Suzzoni, Benjamin",
    title = "{Conformal Defects in Neural Network Field Theories}",
    eprint = "2512.07946",
    archivePrefix = "arXiv",
    primaryClass = "hep-th",
    month = "12",
    year = "2025"
}

@article{Sen:2025vzl,
    author = "Sen, Srimoyee and Vaidya, Varun",
    title = "{Viability of perturbative expansion for quantum field theories on neurons}",
    eprint = "2508.03810",
    archivePrefix = "arXiv",
    primaryClass = "hep-th",
    month = "8",
    year = "2025"
}

@article{Balassa:2025bgt,
    author = "Balassa, Gabor",
    title = "{Addressing the sign problem in Euclidean path integrals with radial basis function neural networks}",
    eprint = "2510.01695",
    archivePrefix = "arXiv",
    primaryClass = "hep-ph",
    doi = "10.1103/74r8-vkm9",
    journal = "Phys. Rev. D",
    volume = "113",
    number = "1",
    pages = "016022",
    year = "2026"
}

@article{Balassa:2025gwh,
    author = "Balassa, Gabor",
    title = "{Neural network approximation of Euclidean path integrals and its application for the $\phi^4$ theory in 1+1 dimensions}",
    eprint = "2509.18785",
    archivePrefix = "arXiv",
    primaryClass = "hep-ph",
    month = "9",
    year = "2025"
}

@article{Balassa:2025biq,
    author = "Balassa, Gabor",
    title = "{On the solution of Euclidean path integrals with neural networks}",
    eprint = "2509.16953",
    archivePrefix = "arXiv",
    primaryClass = "hep-ph",
    month = "9",
    year = "2025"
}

@article{Howard:2024kfd,
    author = "Howard, Jessica N. and Klinger, Marc S. and Maiti, Anindita and Stapleton, Alexander G.",
    title = "{Bayesian RG flow in neural network field theories}",
    eprint = "2405.17538",
    archivePrefix = "arXiv",
    primaryClass = "hep-th",
    doi = "10.21468/SciPostPhysCore.8.1.027",
    journal = "SciPost Phys. Core",
    volume = "8",
    pages = "027",
    year = "2025"
}

@article{Martyn:2022oll,
    author = "Martyn, John M. and Najafi, Khadijeh and Luo, Di",
    title = "{Variational Neural-Network Ansatz for Continuum Quantum Field Theory}",
    eprint = "2212.00782",
    archivePrefix = "arXiv",
    primaryClass = "quant-ph",
    reportNumber = "MIT-CTP/5491",
    doi = "10.1103/PhysRevLett.131.081601",
    journal = "Phys. Rev. Lett.",
    volume = "131",
    number = "8",
    pages = "081601",
    year = "2023"
}

@article{Lee:2025mti,
    author = "Lee, Donghee and Lee, Hye-Sung and Yi, Jaeok",
    title = "{Synaptic field theory for neural networks}",
    eprint = "2503.08827",
    archivePrefix = "arXiv",
    primaryClass = "hep-th",
    doi = "10.1103/bhwt-j2rq",
    journal = "Phys. Rev. D",
    volume = "112",
    number = "3",
    pages = "L031902",
    year = "2025"
}

@article{Ageev:2026qyh,
    author = "Ageev, Dmitry S. and Ageeva, Yulia A.",
    title = "{Neural Network Quantum Field Theory from Transformer Architectures}",
    eprint = "2602.10209",
    archivePrefix = "arXiv",
    primaryClass = "cs.LG",
    month = "2",
    year = "2026"
}

@article{Hashimoto:2024aga,
    author = "Hashimoto, Koji and Hirono, Yuji and Maeda, Jun and Totsuka-Yoshinaka, Jojiro",
    title = "{Neural network representation of quantum systems}",
    eprint = "2403.11420",
    archivePrefix = "arXiv",
    primaryClass = "hep-th",
    reportNumber = "KUNS-2996",
    doi = "10.1088/2632-2153/ad81ac",
    journal = "Mach. Learn. Sci. Tech.",
    volume = "5",
    number = "4",
    pages = "045039",
    year = "2024"
}

@article{Huang:2025ipy,
    author = "Huang, Guojun and Zhou, Kai",
    title = "{The neural networks with tensor weights and emergent fermionic Wick rules in the large-width limit}",
    eprint = "2507.05303",
    archivePrefix = "arXiv",
    primaryClass = "hep-th",
    doi = "10.1016/j.physletb.2025.140146",
    journal = "Phys. Lett. B",
    volume = "873",
    pages = "140146",
    year = "2026"
}

@book{Rychkov:2016iqz,
    author = "Rychkov, Slava",
    title = "{EPFL Lectures on Conformal Field Theory in D{\ensuremath{>}}= 3 Dimensions}",
    eprint = "1601.05000",
    archivePrefix = "arXiv",
    primaryClass = "hep-th",
    reportNumber = "CERN-TH-2016-012",
    doi = "10.1007/978-3-319-43626-5",
    isbn = "978-3-319-43625-8, 978-3-319-43626-5",
    series = "SpringerBriefs in Physics",
    month = "1",
    year = "2016"
}

@article{Dolan:2000ut,
    author = "Dolan, F. A. and Osborn, H.",
    title = "{Conformal four point functions and the operator product expansion}",
    eprint = "hep-th/0011040",
    archivePrefix = "arXiv",
    reportNumber = "DAMTP-2000-125",
    doi = "10.1016/S0550-3213(01)00013-X",
    journal = "Nucl. Phys. B",
    volume = "599",
    pages = "459--496",
    year = "2001"
}

@article{Jakobczyk:1987zx,
    author = "Jakobczyk, L. and Strocchi, F.",
    title = "{KREIN STRUCTURES FOR WIGHTMAN AND SCHWINGER FUNCTIONS}",
    reportNumber = "SISSA-72/87/FM",
    doi = "10.1063/1.527965",
    journal = "J. Math. Phys.",
    volume = "29",
    pages = "1231",
    year = "1988"
}

@article{Jakobczyk:1984ip,
    author = "Jakobczyk, L.",
    title = "{BORCHERS ALGEBRA FORMULATION OF AN INDEFINITE INNER PRODUCT QUANTUM FIELD THEORY}",
    doi = "10.1063/1.526165",
    journal = "J. Math. Phys.",
    volume = "25",
    pages = "617--622",
    year = "1984"
}

@article{Jakobczyk:1987zw,
    author = "Jakobczyk, L. and Strocchi, F.",
    title = "{Euclidean Formulation of Quantum Field Theory Without Positivity}",
    reportNumber = "SISSA-59-87-FM",
    doi = "10.1007/BF01218343",
    journal = "Commun. Math. Phys.",
    volume = "119",
    pages = "529",
    year = "1988"
}

@article{Dobrev:1975ru,
    author = "Dobrev, V. K. and Petkova, V. B. and Petrova, S. G. and Todorov, I. T.",
    title = "{Dynamical Derivation of Vacuum Operator Product Expansion in Euclidean Conformal Quantum Field Theory}",
    reportNumber = "Print-75-0848 (IAS, PRINCETON)",
    doi = "10.1103/PhysRevD.13.887",
    journal = "Phys. Rev. D",
    volume = "13",
    pages = "887",
    year = "1976"
}

@article{Fortin:2016dlj,
    author = "Fortin, Jean-Fran{\c{c}}ois and Skiba, Witold",
    title = "{Conformal Differential Operator in Embedding Space and its Applications}",
    eprint = "1612.08672",
    archivePrefix = "arXiv",
    primaryClass = "hep-th",
    doi = "10.1007/JHEP07(2019)093",
    journal = "JHEP",
    volume = "07",
    pages = "093",
    year = "2019"
}

@inbook{Halvorson:2006wj,
    author = "Halvorson, Hans and Muger, Michael",
    editor = "Butterfield, Jeremy and Earman, John",
    title = "{Algebraic quantum field theory}",
    booktitle = "{Philosophy of physics}",
    eprint = "math-ph/0602036",
    archivePrefix = "arXiv",
    doi = "10.1016/B978-044451560-5/50011-7",
    pages = "731--864",
    year = "2007"
}

@book{Haag:1996hvx,
    author = "Haag, Rudolf",
    title = "{Local Quantum Physics}",
    doi = "10.1007/978-3-642-61458-3",
    isbn = "978-3-540-61049-6, 978-3-642-61458-3",
    publisher = "Springer",
    address = "Berlin",
    series = "Theoretical and Mathematical Physics",
    year = "1996"
}

@article{Osterwalder:1974tc,
    author = "Osterwalder, Konrad and Schrader, Robert",
    title = "{Axioms for Euclidean Green's Functions. 2.}",
    reportNumber = "Print-74-1480 (HARVARD)",
    doi = "10.1007/BF01608978",
    journal = "Commun. Math. Phys.",
    volume = "42",
    pages = "281",
    year = "1975"
}

@book{Glimm:1987ng,
    author = "Glimm, J. and Jaffe, Arthur M.",
    title = "{QUANTUM PHYSICS. A FUNCTIONAL INTEGRAL POINT OF VIEW}",
    year = "1987"
}

@article{osti_4606723,
  author       = {Wightman, A S and Garding, L},
  title        = {FIELDS AS OPERATOR-VALUED DISTRIBUTIONS IN RELATIVISTIC QUANTUM THEORY},
  url          = {https://www.osti.gov/biblio/4606723},
  journal      = {Arkiv Fys.  },
  volume       = {Vol: 28},
  place        = {Country unknown/Code not available},
  year         = {1965},
  month        = {01}}

@article{Mack:1975je,
    author = "Mack, G.",
    title = "{All unitary ray representations of the conformal group SU(2,2) with positive energy}",
    reportNumber = "DESY-75-50",
    doi = "10.1007/BF01613145",
    journal = "Commun. Math. Phys.",
    volume = "55",
    pages = "1",
    year = "1977"
}

@article{Ferko:2026ken,
    author = "Ferko, Christian and Halverson, James and Jejjala, Vishnu and Robinson, Brandon",
    title = "{Topological Effects in Neural Network Field Theory}",
    eprint = "2604.02313",
    archivePrefix = "arXiv",
    primaryClass = "hep-th",
    month = "4",
    year = "2026"
}

@article{Ferko:2026kkm,
    author = "Ferko, Christian and Frank, Samuel and Halverson, James and Jejjala, Vishnu",
    title = "{Anomalies in Neural Network Field Theory}",
    eprint = "2605.12488",
    archivePrefix = "arXiv",
    primaryClass = "hep-th",
    month = "5",
    year = "2026"
}

@article{Hislop:1988yg,
    author = "Hislop, P. D.",
    title = "{CONFORMAL COVARIANCE, MODULAR STRUCTURE, AND DUALITY FOR LOCAL ALGEBRAS IN FREE MASSLESS QUANTUM FIELD THEORIES}",
    doi = "10.1016/0003-4916(88)90044-9",
    journal = "Annals Phys.",
    volume = "185",
    pages = "193--230",
    year = "1988"
}

@article{Hislop:1981uh,
    author = "Hislop, Peter D. and Longo, Roberto",
    title = "{Modular Structure of the Local Algebras Associated With the Free Massless Scalar Field Theory}",
    reportNumber = "UCB-PTH-81/4",
    doi = "10.1007/BF01208372",
    journal = "Commun. Math. Phys.",
    volume = "84",
    pages = "71",
    year = "1982"
}

@article{Buchholz:2019rem,
    author = "Buchholz, Detlev and Fredenhagen, Klaus",
    title = "{A C*-algebraic Approach to Interacting Quantum Field Theories [doi: 10.1007/s00220-020-03700-9]}",
    eprint = "1902.06062",
    archivePrefix = "arXiv",
    primaryClass = "math-ph",
    doi = "10.1007/s00220-021-04213-9",
    journal = "Commun. Math. Phys.",
    volume = "377",
    number = "2",
    pages = "947--969",
    year = "2020"
}

@article{Brunetti:2021wev,
    author = {Brunetti, Romeo and D{\"u}tsch, Michael and Fredenhagen, Klaus and Rejzner, Kasia},
    title = "{C*-algebraic approach to interacting quantum field theory: inclusion of Fermi fields}",
    eprint = "2103.05740",
    archivePrefix = "arXiv",
    primaryClass = "math-ph",
    doi = "10.1007/s11005-022-01590-7",
    journal = "Lett. Math. Phys.",
    volume = "112",
    number = "5",
    pages = "101",
    year = "2022"
}

@inbook{Fredenhagen:2015utr,
    author = "Fredenhagen, Klaus",
    editor = "Brunetti, Romeo and Dappiaggi, Claudio and Fredenhagen, Klaus and Yngvason, Jakob",
    title = "{An Introduction to Algebraic Quantum Field Theory}",
    booktitle = "{Advances in Algebraic Quantum Field Theory}",
    doi = "10.1007/978-3-319-21353-8_1",
    pages = "1--30",
    year = "2015"
}

@article{Hofmann:1995ue,
    author = "Hofmann, G.",
    title = "{On GNS representations on indefinite inner product spaces. 1. The Structure of the representation space}",
    reportNumber = "LEIPZIG-23-1995",
    doi = "10.1007/s002200050270",
    journal = "Commun. Math. Phys.",
    volume = "191",
    pages = "299--323",
    year = "1998"
}

@article{Hofmann:1993dm,
    author = "Hofmann, G.",
    title = "{An Explicit realization of a GNS representation in a Krein space}",
    doi = "10.2977/prims/1195167273",
    journal = "Publ. Res. Inst. Math. Sci. Kyoto",
    volume = "29",
    pages = "267--287",
    year = "1993"
}

@article{Dogra:2026hfa,
    author = "Dogra, Manas and Halverson, James and Naskar, Joydeep",
    title = "{Spinning Conformal Correlators from Neural Networks}",
    eprint = "2608.15001",
    archivePrefix = "arXiv",
    primaryClass = "hep-th",
    month = "8",
    year = "2026"
}

@article{Jefferson:2026nio,
    author = "Jefferson, Ro and Ramakrishnan, Shradha",
    title = "{Deep neural networks as lattice gauge theories}",
    eprint = "2608.19331",
    archivePrefix = "arXiv",
    primaryClass = "hep-th",
    month = "8",
    year = "2026"
}

@inproceedings{Ferko:2026ukw,
    author = "Ferko, Christian and Jejjala, Vishnu and Robinson, Brandon",
    title = "{A Tale of Two Compact Bosons}",
    eprint = "2608.06376",
    archivePrefix = "arXiv",
    primaryClass = "hep-th",
    month = "8",
    year = "2026"
}

@inproceedings{Halverson:2026pmb,
    author = "Halverson, James",
    title = "{Pre-Strings Lectures on Artificial Intelligence}",
    eprint = "2607.02905",
    archivePrefix = "arXiv",
    primaryClass = "hep-th",
    month = "7",
    year = "2026"
}

@article{Balassa:2026lil,
    author = "Balassa, Gabor",
    title = "{Approximating Grassmann valued path integrals with radial basis function neural networks}",
    eprint = "2608.10458",
    archivePrefix = "arXiv",
    primaryClass = "hep-lat",
    doi = "10.1103/c918-1gq1",
    journal = "Phys. Rev. D",
    volume = "114",
    number = "3",
    pages = "036024",
    year = "2026"
}

@article{Harvey:2026iqd,
    author = "Harvey, Thomas R.",
    title = "{What Neural Network Field Theory Can and Cannot Realise on a Computer}",
    eprint = "2608.21523",
    archivePrefix = "arXiv",
    primaryClass = "hep-th",
    month = "8",
    year = "2026"
}

@article{Ribault:2026bbu,
    author = "Ribault, Sylvain",
    title = "{Conformal correlator systems}",
    eprint = "2609.31237",
    archivePrefix = "arXiv",
    primaryClass = "hep-th",
    month = "9",
    year = "2026"
}

\end{document}